%% file: 7281.tex
\documentclass[oldversion]{aa}
\usepackage{graphicx}
\usepackage{natbib}
\usepackage{amsmath}
\usepackage{txfonts}
\usepackage{lscape,longtable,array}

\bibpunct{(}{)}{;}{a}{}{,}

\newcommand{\kms}{km\,s$^{-1}$}
\newcommand{\dif}{\mathrm{d}}
\newcommand{\map}{M_\mathrm{ap}}

\newcommand{\im}{\mathrm{i}}
\newcommand{\ex}{\mathrm{e}}
\newcommand{\h}{h_{70}}
\newcommand{\hm}{h_{70}^{-1}}

\begin{document}
\title{BLOX: The Bonn Lensing, Optical, and X-ray selected \\
  galaxy clusters} 
\subtitle{I. Cluster catalog construction\thanks{Based on
  observations carried out at the European Southern Observatory, La
  Silla, Chile under program Nos. 170.A-0789, 70.A-0529, 71.A-0110,
  072.A-0061, 073.A-0050.}, \thanks{The cluster catalogs are available
  in electronic format at the CDS via anonymous ftp to
  \texttt{cdsarc.u-strasbg.fr (130.79.128.5)} or via
  \texttt{http://cdsweb.u-strasbg/cgi-bin/qcat?J/A+A/VOL/PAGE}}}
\author{J.\,P.~Dietrich\inst{1,2}
  \and
  T.~Erben\inst{1}
  \and
  G.~Lamer\inst{3}
  \and
  P.~Schneider\inst{1}
  \and
  A.~Schwope\inst{3}
  \and
  J.~Hartlap\inst{1}
  \and
  M.~Maturi\inst{4}
}

\offprints{J.~P.~Dietrich}

\institute{Argelander-Institut f\"ur Astronomie, University of Bonn,
  Auf dem H\"ugel 71, 53121 Bonn, Germany 
  \\\email{jdietric@eso.org} 
  \and
  ESO, Karl-Schwarzschild-Str. 2, 85748 Garching b. M\"unchen, Germany
  \and
  Astrophysikalisches Institut Potsdam, An der Sternwarte 14, 14482
  Potsdam, Germany 
  \and 
  Institut f\"ur Theoretische Astrophysik, Albert-Ueberle-Str. 2, 69120
  Heidelberg, Germany}

\date{Received 12 February 2007; Accepted 19 April 2007}

\abstract{The mass function of galaxy clusters is an important
  cosmological probe. Differences in the selection method could
  potentially lead to biases when determining the mass function. 
  From the optical and X-ray data of the XMM-Newton Follow-Up Survey,
  we obtained a sample of galaxy cluster candidates using weak
  gravitational lensing, the optical Postman matched filter method,
  and a search for extended X-ray sources. 
  We developed our weak-lensing search criteria by testing the
  performance of the aperture mass statistic on realistic ray-tracing
  simulations matching our survey parameters and by comparing two
  filter functions. We find that the dominant noise source for our
  survey is shape noise at almost all significance levels and
  that spurious cluster detections due to projections of
  large-scale structures are negligible, except possibly for
    highly significantly detected peaks.
  Our full cluster catalog has 155 cluster candidates, 116 found with
  the Postman matched filter, 59 extended X-ray sources, and 31 shear
  selected potential clusters. Most of these cluster candidates were
  not previously known. The present catalog will be a solid foundation
  for studying possible selection effects in either method.

}
\maketitle

\section{Introduction}
\label{sec:introduction}
Because the dynamical or evolutionary timescale of clusters of galaxies
is not much shorter than the Hubble time, they retain a `memory' of
the initial conditions for structure formation
\citep[e.g.,][]{2001Natur.409...39B}. The population of clusters
evolves with redshift, and this evolution depends on the cosmological
model \citep[e.g.,][]{1996MNRAS.282..263E}; therefore, the redshift
dependence of the cluster abundance has been used as a cosmological
test \citep[e.g.,][]{2003ApJ...590...15V,2004ApJ...609..603H}. The
dependence of the cluster abundance on cosmological parameters can be
obtained either from analytical models \citep{1974ApJ...187..425P} or,
more reliably, from $N$-body simulations
\citep[e.g.,][]{2001MNRAS.321..372J}. What such models do predict is
the abundance of dark matter halos as a function of redshift and mass.

Clusters can be selected by various methods: optical, X-ray emission,
weak lensing \citep[e.g.,][]{1996MNRAS.283..837S}, and -- using future
surveys -- the Sunyaev-Zeldovich effect (SZE). Optical, X-ray, and SZE
selection of clusters depends on the baryonic content of clusters,
which -- compared to the predicted dark matter density -- is a minor
fraction of the clusters' constituents. Optical selection depends on
the star formation history, and X-ray detection requires a hot
intracluster medium (ICM). \citet{2002MNRAS.337.1269W} predict that up
to 20\% of all weak lensing clusters have not heated their ICM to a
level detectable with current X-ray missions. Searching for clusters
with SZE is a very promising method but has yet to produce first
samples.

Common to all methods except X-ray selection is that they are prone to
projections along the line-of-sight. Spectroscopy is an essential tool
in distinguishing real clusters from chance projections on the sky. X-ray
emission, which depends on the square of the local gas density, is not
easily affected by line-of-sight projections but is susceptible to
other sorts of biases, e.g., heating of the ICM in mergers.

All four selection methods are sensitive in different redshift
regimes. Optical and X-ray searches depend on the luminosity distance
of clusters. SZE is nearly redshift independent. Weak lensing surveys
typically cover the redshift range of 0.15--0.7. Clearly, no cluster
selection method is ideal, and understanding their biases and
limitations is important for precision cosmology using clusters.

Several galaxy clusters have already been found using weak
gravitational lensing in recent years: in the FORS1 cosmic shear
survey \citep{2001A&A...368..766M,2005A&A...442...43H}, one mass peak
clearly coincides with an overdensity of galaxies.
\citet{2005A&A...440..453D} found a cluster in the background of the
super-cluster system A~222/223. \citet{2006ApJ...643..128W}, have
published 6 new clusters detected with weak lensing from their Deep
Lens Survey. These examples clearly demonstrate that this method of
cluster detection in fact works.

The selection of clusters of galaxies by weak lensing, however, also
faces significant methodological challenges. Even before the first
spectroscopically confirmed weak-lensing detected cluster was reported
\citep{2001ApJ...557L..89W}, \citet{2000A&A...355...23E} reported a
highly significant tangential alignment of galaxies around an empty
spot on the sky \citep[see also][]{2006A&A...454...37V}. Two more of
these \emph{dark clumps} have been reported in the literature
\citep[][but see also
\citealt{2003A&A...410...45E}]{2000ApJ...539L...5U,2002A&A...388...68M}.
Recently \citet{2007A&A...462..875S} have published a catalog of 158
shear selected peaks, 72 of them associated with bright galaxy
concentrations. Of course, the reality of these dark clump detections
has to be considered with caution, given that even one of them would
have a profound impact on our understanding of the evolution of dark
and baryonic matter in the Universe.

Several theoretical studies have recently shed some light on the
problem of dark clump detection in weak lensing surveys. Among them
are
\citet[][H04]{2004MNRAS.350..893H}\defcitealias{2004MNRAS.350..893H}{H04}
and
\citet[][HS05]{2005ApJ...624...59H}\defcitealias{2005ApJ...624...59H}{HS05}
who have both used ray-tracing simulations through $N$-body
simulations to study the \emph{efficiency}, also called \emph{purity}
by other authors, and \emph{completeness} of the detection of clusters
of galaxies in weak lensing surveys. An important result of
\citetalias{2005ApJ...624...59H} is that the efficiency, even in the
limiting case of no intrinsic galaxy ellipticity, does not increase
beyond $85\%$. The remaining $15\%$ of shear selected peaks are due to
projections of the large-scale structure along the line of sight and
will be seen as dark clumps. These could very well account for the
dark clumps studied so far in detail in the literature. The efficiency
naturally drops further if more realistic noise caused by the
ellipticity of the background galaxies is assumed. The completeness
was studied in more depth by \citetalias{2004MNRAS.350..893H}, who
find that, even with low significance thresholds in the selection
process shear, selected samples will be incomplete, except at the
highest masses.

In this paper we describe a search for galaxy clusters in the public
XMM-Newton Follow-Up Survey \citep{2006A&A...449..837D} and our
private extension to the survey. We combine the results of three
independent selection methods using the aperture mass statistic
\citep{1996MNRAS.283..837S} for weak lensing selection, the optical
matched filter algorithm \citep[hereinafter
P96]{1996AJ....111..615P}\defcitealias{1996AJ....111..615P}{P96} for
optical selection of galaxy clusters, and a search for extended X-ray
emission in the XMM-Newton data on our survey fields. The combination
allows us to dig deeper into the mass function than we could do with
weak lensing selection alone.

This paper is organized as follows:
Section~\ref{sec:xmm-newton-follow} briefly describes the XMM-Newton
Follow-Up Survey, the optical data, and their reduction.
Section~\ref{sec:x-ray-detection} gives an overview of the X-ray data,
their reduction, and then describes how we detected galaxy cluster
candidates as extended X-ray sources. In
Sect.~\ref{sec:optic-match-filt} we present our implementation of the
matched filter technique of \citetalias{1996AJ....111..615P} and the
resulting catalog of cluster candidates. We briefly summarize the
aperture mass methods \citep{1996MNRAS.283..837S} in
Section~\ref{sec:weak-lens-detect} and develop our selection criteria
used for cluster detection later in that section based on realistic
ray-tracing simulations. We discuss and summarize our findings in
Sect.~\ref{sec:summary}. The catalogs of cluster candidates are
available in electronic format at the CDS.

Throughout this work we assume an $\Omega_\mathrm{m}=0.3$,
$\Omega_\Lambda = 0.7$, $H_0 = 70\,\h$\,\kms\ cosmology and use
standard lensing notation \citep[e.g.,][]{2001Phys.Rep..340..291B}.

\section{The XMM-Newton Follow-Up Survey}
\label{sec:xmm-newton-follow}
The XMM-Newton Follow-Up Survey (XFS) consists of two parts, a public
and a private one. Both were conducted with the Wide-field Imager
(WFI) at the ESO/MPG-2.2\,m telescope on La Silla, Chile. The survey
provides optical imaging on fields for which deep, public XMM-Newton
exposures exist.

\subsection{Public and private survey}
\label{sec:publ-priv-surv}
The public survey (ESO Program Id. 170.A-0789, P.I. J. Krautter as
chairman of the ESO working group for public surveys) was carried out
in the framework of the ESO Imaging Survey (EIS) as a collaboration
between ESO, the XMM-Newton Survey Science Centre (SSC), and a group
at the Institut f\"ur Astrophysik und Extraterrestrische Forschung
(IAEF) at the University of Bonn. The aim of the public survey, whose
observations were concluded in November 2005, was to provide optical
counterparts to serendipitously detected X-ray sources in the southern
hemisphere. To provide data for a minimum spectral discrimination and
photometric redshift estimates, the public survey was carried out in
the B-, V-, R-, and I-passbands down to a limiting magnitude of
$25$\,mag in the AB system in all bands. All data of the public survey
taken before October 16, 2003 were reduced, calibrated, and publically
released in July 2005 \citep{2006A&A...449..837D}. The total public
survey comprises 15 WFI fields. Out of these, 4 are galactic fields
and thus unsuitable for galaxy cluster searches. One field is a
mispointing without X-ray data and high galactic absorption. The
remaining 10 fields cover about $\sim2.8$\,sq.\,deg. in BVRI.

The private extension of the survey (Program Ids. 70.A-0529,
71.A-0110, 072.A-0061, 073.A-0050, P.I. P.~Schneider) with the goal of
a weak lensing search for galaxy clusters has been conducted as a
collaboration between the IAEF and the Astrophysikalisches Institut
Potsdam (AIP). Originally targeted to observe 17 additional fields in
B- and R-band and to be finalized by October 2003, observations were
concluded only in September 2006 due to weather and scheduling
problems. Out of the 14 additional fields observed in R-band, our
primary band for the lensing analysis, 13 are used for this work. This
corresponds to all data obtained until September 30, 2005. B-band
observations, proposed to allow for a red-sequence cluster search
\citep{2000AJ....120.2148G}, are available for 9 of these fields.

A cluster survey carried out on these fields is obviously not an
unbiased survey since it includes 5 XMM-Newton observations that
targeted known galaxy clusters. The impact on the total number of
clusters detected, however, is small, at least for the X-ray and
optically detected clusters. The influence on the weak lensing sample
is greater because weak lensing requires clusters to be fairly massive
to be detectable with 2\,m class telescopes. Previously known clusters
are likely to be relatively massive and influence the smaller sample
of weak lensing detected clusters more strongly.

\subsection{Optical data reduction}
\label{sec:optic-data-reduct}
We reduced the optical XFS data using the publically available GaBoDS
pipeline \citep{2005AN....326..432E} with the Guide Star Catalog version
2.2 (GSC-2.2) as the astrometric reference and the
\citet{2000PASP..112..925S} catalog for photometric calibration. A
subset of our reduced data was compared against the publically
released XFS data \citep{2006A&A...449..837D} and found to be in very
good agreement with this independent reduction.

We would like to point out that we applied the fringing removal
procedure to all our R-band images to remove the low level fringing
present in WFI R-band. This is a difference to the data released by
\citet{2006A&A...449..837D}. Weight images describing the relative
noise properties of each pixel were created to mask bad pixels or
columns, cosmic rays, and other image defects masked manually, like
satellite tracks and ghost images due to internal reflections.

\input{7281tab1.tex}

Table~\ref{tab:optical_fields} lists all coadded R-band images with
center coordinates of the survey fields used in the present work,
their $5\sigma$ limiting magnitude computed in the Vega system in a
$2\arcsec$ diameter aperture, their seeing, as well as the effective
(unmasked) area used for catalog creation (see
Sect.~\ref{sec:optic-catal-creat}), and the number densities of the
resulting lensing (see Sect.~\ref{sec:weak-lensing-catalog}) and
matched filter (Sect.~\ref{sec:postm-match-filt}) catalogs. 

Two independent observations of the field RBS~0864 are used in the
XFS. The WFI observations in the V- and R-bands done by Schindler et
al. (RBS~0864-N in Table~\ref{tab:optical_fields}) are centered on
coordinates slightly north west of the galaxy cluster and were taken
with a seeing constraint of $1\farcs2$. The re-observation in the B-
and R-bands two years later in the course of the XFS (RBS~0864-S) is
centered on the cluster and was observed with the seeing constraint of
$1\farcs0$ of this survey. However, most data was taken well outside
the specified constraint. Consequently, we used only the RBS~0864-N
pointing for our weak lensing analysis. We will describe the full XFS
data set including BVI passband data in a forthcoming paper (Dietrich
et al., in preparation).

\section{X-ray detection}
\label{sec:x-ray-detection}
The X-ray luminosity of the hot intracluster gas scales with the
square of its density. Thus, X-ray detections of clusters of galaxies
are relatively insensitive to projection effects. The high sensitivity
of XMM-Newton allows us to find clusters out to very high redshifts.
Two of the three most distant and spectroscopically confirmed clusters
were found serendipitously with XMM-Newton. These are at redshift
$z=1.39$ \citep{2005ApJ...623L..85M} and $z=1.45$
\citep{2006ApJ...646L..13S}, the latter located in the public
XFS field \object{LBQS~2212$-$1759}.

\subsection{X-ray data reduction}
\label{sec:x-ray-data}
The archival XMM-Newton data of the XFS were reduced with the latest
version of the Science Analysis System available at that time
(SAS-6.5.0). We briefly describe the standard processing steps
employed to create X-ray source lists from the observation data files.
Appendix~\ref{sec:x-ray-observations} gives an overview of all X-ray
observations considered in the present work. Some fields were
observed more than once. Their data reduction requires additional
steps described at the end of this section.

SAS was used to generate calibrated event lists, exposure maps
describing the spatial variations of the instruments' sensitivity, and
images in five energy bands ranging from $0.1\text{--}0.5$\,keV,
$0.5\text{--}1.0$\,keV, $1.0\text{--}2.0$\,keV,
$2.0\text{--}4.5$\,keV, and $4.5\text{--}12.0$\,keV. We used the
standard SAS flags \texttt{\#XMMEA\_EM} and \texttt{\#XMMEA\_EP} to
filter the event lists when generating the images. For the PN detector
we imposed additional restrictions on the allowed charge patterns. For
the two lowest energy bands we accepted only single events; for the
three highest energy bands only events not depositing charge in more
than two neighboring pixels were kept. We also excluded the energy
range of $7.8\text{--}8.2$\,keV from the PN data to avoid the complex
of Ni-K$\alpha$, Cu-K$\alpha$, and Zn-K$\alpha$ fluorescence lines of
the detector and surrounding structure.

Using the 15 images generated with \texttt{evselect} (5 images per
camera and 3 cameras), the mask images, and the vignetted exposure
maps, \texttt{ebox\-de\-tect} in local mode was run to generate a
first source list. The minimum likelihood for a detection in this step
was set to $5$. The sources were excised from the image by
\texttt{esplinemap} to model the background of the images. The task
\texttt{esplinemap} allows to describe the background either by a
two-component model (vignetted astrophysical and unvignetted particle
background) based on ray-tracing of the instruments or by a 2-d spline
with a user-defined number of nodes. For the XFS the decision about
which approach to use was based on visual inspection of the data.

With these background maps, \texttt{eboxdetect} was run in map mode to
create another source list. The minimum likelihood for source
detection was set to $4$ in this step. This list was given together
with the science images, masks, and background maps to
\texttt{emldetect}, which performs a simultaneous maximum likelihood
multi-source PSF fitting in all energy bands. The free parameters
\texttt{emldetect} fits are the source position, source extent (core
radius of a $\beta$-model), and source count rate in each energy band
and instrument. Derived parameters are the total source count rate,
total likelihood of detection and likelihood of detection per energy
band, likelihood of source extent, and four hardness ratios between
the input energy bands. In our reduction we let \texttt{emldetect} fit
up to two sources to one source position reported by
\texttt{eboxdetect}. The minimum likelihood for a detection was set
to $6$, and the minimum extent likelihood for a source to be
considered as an extended source to $4$.

It should be pointed out that the result of the PSF fitting performed
by \texttt{emldetect} cannot be better than the available PSF models.
Especially at large off-axis distances the EPIC PSF is not very well
known.

For the two deepest combined fields (LBQS~2212$-$1759, 250\,ks,
PG~1115+080, 220\,ks) the source detection process was slightly
modified with respect to the other observations. In deep observations
small inaccuracies of the background fit can lead to many spurious
detections of extended sources. Since in the softest band
($0.2\text{--}0.5$\,keV) both the MOS and PN cameras show spatial
variations of the detector background and since the hardest band
($4.5\text{--}12$\,keV) is strongly dominated by background, only the
3 bands in the range $0.5\text{--}4.5$\,keV were used for source
detection in the two very deep observations.

The source positions in the \texttt{emldetect} catalog have
statistical errors on the order of $1\arcsec\text{--}2\arcsec$, plus a
systematic error due to an uncertainty in the attitude of the
spacecraft, which has the same size. The latter can be corrected by
cross-correlating the X-ray source positions with the more accurate
positions of optical sources. The task \texttt{eposcorr} was used to
carry out this cross-correlation with the R-band catalogs on all
fields. The production of the optical catalogs is described in detail
in Sect.~\ref{sec:optic-catal-creat}.

Areas of the EPIC field-of-view that are dominated by bright extended
emission of the XMM-Newton target were excluded from the survey by
masking out a circular area around the target before starting the
detection process.

Multiple observations of one field are in general taken at different
roll angles and -- since the center of rotation is not the center of
the FOV -- different positions. The first step in combining
observations is thus to bring them all to a common nominal position.
This was done with the SAS task \texttt{attcalc}, which computes sky
coordinates for event files. From these new event lists FITS images,
exposure maps, and masks were created for the individual observations
as described above. Science images and exposure maps were coadded
weighted by the masks of the respective observation. The masks
themselves are combined with logical or. The source extraction
continues on the combined images, exposures maps, and masks as
described above.

\subsection{Catalog of extended X-ray sources}
\label{sec:catalog-extended-x}
Catalogs of cluster candidates were generated from the
\texttt{emldetect} source list. Sources with a detection of likelihood
$> 15$ and extent likelihood $> 4$ were considered as potential
cluster candidates. Any remaining large-scale inhomogeneities in the
background are sometimes detected as spurious sources and the best-fit
model of \texttt{emldetect} is often one whose extent reaches the
maximum value of $20$\,pixels ($80\arcsec$), hence only sources with
an extent $< 20$\,pixels were kept in the final catalog. The extended
X-ray sources found in this way were visually screened and any obvious
artifacts, often due to out-of-time (OOT) events or remaining
background structure, were manually rejected. The optical images were
visually inspected for possible counterparts of extended sources. In
this step extended sources obviously associated with nearby galaxies
were removed from the catalog. Grades were assigned to the quality of
a cluster detection based on visual inspection of the optical and
X-ray images. Cluster candidates with grade ``+'' are obvious real
clusters, often the ones one would select by eye. Extended X-ray
sources graded with ``$\circ$'' are possible clusters but not as
prominent as those graded with ``+''. This grade is assigned to
clusters without a very obvious optical counterpart but a reliable
X-ray detection. In some cases these may be systems at very high
redshift that are just barely visible in the optical images. Clusters
graded ``$-$'' appear to be unreliable in the optical and X-ray
images, but were not rejected as obvious spurious sources based on
their visual impression in the X-ray images.

The full X-ray cluster catalog is available in electronic format from
the CDS. Cluster candidates in this catalog follow the naming scheme
BLOX~JHHMM.m+DDMM.m, where BLOX is the IAU registered acronym for
``Bonn Lensing, Optical, X-ray'' detected cluster candidates. In the
following we give comments on individual fields when appropriate.
\begin{itemize}
\item \object{RX J0505.3$-$2849} -- Two extended X-ray peaks are each
  found on both previously known RX~J clusters in this field. In the
  case of \object{RX~J0505.3$-$2849}, we only list one; the other is
  probably a confusion with a double point source;
\item \object{RBS~0864} -- Two reductions of this field were done. One
  reduction excluded the target cluster from the analysis,
  while the other one was made with the cluster remaining in the data.
  A large number of spurious detections associated with OOT events
  were manually rejected.
\item \object{MS~1054.4$-$0321} -- Two extended X-ray sources were
 detected on the target cluster at about the same distance from the
 optical center of the cluster. We list both X-ray sources in the
 catalog. 
\item \object{HE~1104$-$1805} -- The only extended X-ray source found
 in this field coincides with a bright star.
\item \object{LBQS~1228+1116} -- The X-ray data on this field is
 strongly affected by background flares. No clusters were found in
 the remaining shallow data.
\item \object{MKW~9} -- The target cluster dominates the center and
 several extended sources are detected in the cluster region, some of
 them on bright and large galaxies.
\item \object{QSO B1246$-$057} -- The calibration problem that
  prevented the X-ray data on this field from being included in the
  public data release \citep{2006A&A...449..837D} were solved and the
  field could be included in the cluster search.
\item \object{A~1882} -- The X-ray image of this cluster shows three very
  extended sources centered around the nominal position of the cluster
  of \citet{1958ApJS....3..211A}. These regions of extended X-ray
  emission were excluded from the analysis.
\item \object{NGC~7252} -- No extended targets were found in this field.
\end{itemize}

\section{Optical matched filter detection}
\label{sec:optic-match-filt}
Clusters of galaxies can be optically selected in a multitude of ways,
either from one passband alone or by combining color information from
two or more passbands with the positional information on galaxies used
in all methods. A review of a large number of optical detection
methods has recently been published by \citet{2006astro.ph..1195G}.

For this work we chose the matched filter detection algorithm of
\citet{1996AJ....111..615P}. The \citetalias{1996AJ....111..615P}
method was selected because it is well-tested and efficient, it works on
single passband catalogs, and can thus be used for the entire area
of the XFS. More elaborate detection schemes using multi-color
information will be employed on XFS data in subsequent work (Dietrich
et al. in preparation).

\subsection{Optical catalog creation}
\label{sec:optic-catal-creat}
The starting point for any optical cluster-search method is a catalog
of galaxies. It is thus only prudent to discuss the creation of such
catalogs in some detail before turning to a short description of the
matched filter algorithm and how it was implemented for this survey.

Unfortunately, WFI images are especially prone to internal
reflections, producing prominent reflection rings around all saturated
stars with blooming in the core. The brightest stars show more than
one reflection ring, with increasing sizes and offsets from the
position of the source. Additionally, extended halos and diffraction
spikes occur around bright sources. The catalog creation tries to mask
regions affected by these problems and do so as automatically as
possible, but it still requires a large amount of manual masking.

The catalog production uses \texttt{SExtractor}
\citep{1996A&AS..117..393B} and starts with a very low signal-to-noise
ratio (SNR) catalog. The weight images produced by the GaBoDS pipeline
are used in all steps of the catalog creation. The sole purpose of
this first catalog is to identify regions that should be masked.
Masked regions will be passed on to \texttt{SExtractor} in a FLAG
image.

All objects of the initial catalog are put into cells of a grid whose
size is chosen such that the average number of objects per cell is
$2$. This grid is smoothed with a Gaussian kernel with an FWHM of
$1.4$ pixels. We call the smoothed array $\mathcal{S}$. Automatically
adjusting the grid size such that a fixed number of galaxies per cell
is reached on average allows one to keep the size of the Gaussian,
which is necessary to reach the desired SNR in the object density
distribution, as fixed in the program. The main advantage of this
approach is that the values of the kernel can be stored in a matrix
(fixed to a $5\times 5$ array in the program). The convolution of the
density grid with the kernel matrix is computationally much faster
than re-evaluating the Gaussian kernel at each grid cell. In the next
step the dynamic range $\mathcal{S}$ is limited. Every pixel with a
value $> 1.5$ is set to $2$. Every pixel with a value $< 1$ is set to
zero; these pixels will be masked because they are either in very low
SNR parts of the image, such as the edges, or they are covered by
extended objects, such as large foreground galaxies or very bright
stars. The resulting array is called $\mathcal{D}$.

Additionally, any rapid change of object density in $\mathcal{D}$,
such as seen at the edges of reflection rings, is detected with a
Sobel edge detection, i.e., an array containing the absolute values of
the gradient of the array $\mathcal{D}$. The gradient computation
is implemented as a convolution of $\mathcal{D}$ with the two $3
\times 3$ convolution kernels that correspond to finite second-order,
two-sided differentiation. Every pixel exceeding a threshold in
$\mathcal{D}$ is flagged in the output image.

Finally, the FLAG array, containing only values of $1$ for
pixels to be flagged and $0$ for all other pixels, is smoothed again
with the $1.4$-pixel FWHM Gaussian to account for the fact
that the initial smoothing shrinks the areas not covered by objects.
Every pixel with a non-zero flag value will be flagged in the output
FLAG image, which is expanded to the size of the original
WFI image from which the catalog was created.

FLAG images created through this procedure reliably mask extended
objects, bright stars, and the most prominent reflection rings.
Exceptionally empty regions on the sky are only rarely masked
erroneously. However, fainter reflection rings and stars of
intermediate magnitude must still be masked by hand. Files describing
the regions masked manually -- either by circles or polygons for more
complex shapes -- must be generated by the user with tools such as the
image viewer DS9. From them and the automatically generated FLAG
image, the final FLAG image used in further steps in the catalog
production is created.

The image seeing is determined from a high-SNR catalog. A histogram of
the FWHM of all objects with $0\farcs3 < \text{FWHM} < 3\farcs0$ is
created and the image seeing is set equal to the FWHM of the bin with
the most objects. The seeing is used as input for
\texttt{SExtractor}'s star-galaxy classifier \texttt{CLASS\_STAR}.
Since we are only interested in galaxies and not in stars, reliably
separating them in the science-grade catalogs is important for galaxy
cluster searches.

Finally, two science-grade catalogs are created with
\texttt{SExtractor}. Their only difference is the filter with which
the detection image is convolved. The catalog for the optical matched
filter search is made using a Gaussian kernel with an FWHM of $4$
pixels. This kernel ensures that relatively few spurious detections of
faint objects are made, but it has a lower completeness at the low SNR
end with a tendency to miss very small objects. Another catalog with
\texttt{SExtractor}'s default filter -- a $3\times 3$ pixel pyramidal
kernel -- is created as the starting point for the weak-lensing
catalog creation (see Sect.~\ref{sec:weak-lens-detect}).

\subsection{The Postman matched filter catalog}
\label{sec:postm-match-filt}
The Postman matched filter algorithm is described in detail in
\citetalias{1996AJ....111..615P}. The main features of this algorithm
separating it from other single-band cluster detection schemes are (1)
it uses photometric and not only positional information (2) the
contrast of structures that approximate the filter shape with the
background is maximized, (3) redshift and richness estimates of
cluster candidates are produced as a byproduct, (4) the algorithm
performs well over a wide range of redshifts. The main disadvantage is
that a particular radial profile and luminosity function must be
assumed. Clusters deviating from the expected profile will be detected
only at lower significances or suppressed. Our implementation follows
the description of \citetalias{1996AJ....111..615P} and \citet[][O99
hereinafter]{1999A&A...345..681O}\defcitealias{1999A&A...345..681O}{O99}
with some modifications of the selection parameters as outlined below.

A series of programs and shell scripts is used to go from the object
catalogs described in Sect.~\ref{sec:optic-catal-creat} to a catalog
of cluster candidates. The input catalog is filtered such that only
objects with a high probability of being galaxies are kept. To
this end only objects with a \texttt{CLASS\_STAR} value $<0.5$ are
kept. At the faint end the catalog is cut at a magnitude $10\sigma$
above the sky background as measured in a $2\arcsec$ aperture to
ensure a high completeness of the catalog. Objects brighter than
$17$\,mag are also filtered. These are often bright stars, which are
not correctly identified as such by \texttt{SExtractor} and would
cause a serious contamination of the input catalog. If any of these
objects with $R<17$ are bright, nearby cluster members, the matched
filter signal of the cluster is decreased. However, the fainter
cluster galaxies usually still lead to a significant detection of the
galaxy cluster but with a redshift bias. This bias is introduced
because omitting the bright cluster galaxies modifies the luminosity
function. This redshift bias introduced by the cut on the bright
galaxies is marginal, at least for the redshift range considered here.

Likelihood maps $S(x, y)$ are computed on a grid and saved as
\texttt{FITS} images for a series of redshifts. We compute $S$ with a
spacing in redshift of $\Delta z  = 0.1$, starting at $z = 
0.1$ and up to the last output grid in which the $m^*$ of the fiducial
cluster model shifted to that redshift is not fainter than the
limiting magnitude of the input catalog. For the XFS R-band
observations, this is typically up to $z=0.9$. Following the
prescription of \citetalias{1999A&A...345..681O}, we assume a fiducial
cluster model at $z=0.6$ with a Schechter luminosity function with
$M_R^* = -21.63$\,mag, faint end slope $\alpha = -1.1$, and a
Hubble profile with physical core radius $R_\mathrm{c} = 
0.1\,\h$\,Mpc integrated out to the cutoff radius $r_\infty = 10
R_\mathrm{c}$. We keep the pixel size of the output grid constant at
$0.5$\,pix/$R_\mathrm{c}$ at $z=0.6$ but vary the area required for a
detection in the output images with redshift. The $k$-correction
needed for the redshift dependence of $m^*$ is computed for an
elliptical galaxy with no evolution.

Having created the likelihood maps, we are now faced with the problem
of identifying clusters of galaxies in them; i.e., we have to decide
which peaks in the likelihood maps are reliable candidates for galaxy
clusters. Peaks are detected with \texttt{SExtractor}. The main
\texttt{SExtractor} detection parameters were adopted from
\citetalias{1999A&A...345..681O} and are set as follows:
\begin{itemize}
\item The minimum area for a detection \texttt{MIN\_AREA} scales with
  the redshift and corresponds to $\pi r_\mathrm{c}^2$, where
  $r_\mathrm{c}$ is the size of core radius $R_\mathrm{c}$ projected
  on the sky;
\item The detection threshold \texttt{DETECT\_THRESH} is set to $2$;
\item No deblending of peaks is performed (\texttt{DEBLEND\_MINCONT}$
 = 1$), so that all contiguous pixels above the detection
  threshold are counted as one cluster candidate.
\item A global background is estimated from the image.
\end{itemize}

Catalogs from the individual output grids at different redshifts are
then matched by position using the LDAC program \texttt{associate}.
Peaks present in at least $3$ output grids with a minimum significance
of at least $3.5\sigma$ in one of them are kept as reliable cluster
candidates. These parameters are slightly different from the ones
adopted by \citetalias{1999A&A...345..681O}, who used a threshold of
$3\sigma$, a detection in at least $4$ output grids, and a minimum
value of the richness parameter $\Lambda_\mathrm{cl}$. The reason for
different selection criteria are the different redshift regions of
interest in the study of \citetalias{1999A&A...345..681O} and the work
presented here. While \citetalias{1999A&A...345..681O} are chiefly
interested in high redshift clusters, our search mainly aims for
intermediate redshift clusters accessible with weak lensing, i.e.,
most of these clusters will be at redshifts $z = 0.2\ldots 0.3$.
Several of the criteria adopted by \citetalias{1999A&A...345..681O}
are biased against the search for clusters at intermediate redshifts.

The significance of a cluster signal drops sharply once we look at
output grids at redshifts higher than the cluster's redshift because
the power law cutoff of the luminosity filter strongly suppresses
signals from lower redshifts. For clusters at lower redshifts, the
number of output grids at redshifts lower or equal to the cluster
redshift is small. The requirement of
\citetalias{1999A&A...345..681O} that a cluster candidate be detected
in at least $4$ output grids is thus heavily biased against the detection
of clusters at the redshift interval we are especially interested in.
We therefore relaxe the requirement on the number of output grids in
which a cluster must be detected to $3$, and have the lowest redshift
output grid at $z=0.1$ instead of $z=0.2$. To compensate for the
higher number of spurious detections caused by this less stringent
cut, the minimum significance to be reached in at least one output
grid is increased from $3$ to $3.5$.

We do not make any cuts on the richness parameter
$\Lambda_\mathrm{cl}$. Clusters at higher redshifts need to be much
more luminous than clusters at lower redshifts to be detectable.
Consequently, the average richness parameter of clusters at higher
redshift is much larger than that of lower redshift clusters. Imposing
a minimum value for $\Lambda_\mathrm{cl}$ would create a strong bias
against low and intermediate redshift systems.

The selection criteria adopted by \citetalias{1999A&A...345..681O}
have proven to be very successful by a high rate of spectroscopically
confirmed clusters found in the EIS wide survey \citep[e.g.,
][]{2002A&A...388....1H,2003A&A...409..439O,2002A&A...394....1B} and
we are thus confident that our slightly modified criteria are also
successful. However, we have to note that the simulations on which
these criteria were developed are not able to predict the number of
spurious detections. \citetalias{1999A&A...345..681O} attempted to
simulate a pure background population of galaxies by randomizing the
position of galaxies while keeping their magnitude fixed. This
randomized population has a much smoother distribution than the field
population of galaxies. The reason is that the distribution of field
galaxies is not a random field with white noise, but instead shows
clustering and correlation.

More recently,
\citet[][O07]{2007A&A...461...81O}\defcitealias{2007A&A...461...81O}{O07}
have used a more realistic approach by creating backgrounds that are
correlated within bins of 1~mag following the prescription of
\citet{1978AJ.....83..845S}. Their selection parameters are a minimum
significance of $3.5\sigma$, minimum area of $\pi r_\mathrm{c}^2$, and
number of redshift slices $>1$, which are very similar to ours.

We applied the matched filter algorithm to R-band catalogs from 23
fields of the XFS. The resulting catalog is available in electronic
format from the CDS. It lists $116$ candidate clusters. Thirteen of
the systems in this catalog have been previously found by other
authors and have either spectroscopic or photometric redshift
information. All cluster candidates were visually inspected and graded
as for to the grades given to the X-ray detected clusters. Cluster
candidates with grade ``+'' are obvious real clusters, often the ones
one would select by eye, or they have and extended X-ray source as a
counterpart. Matched-filter peaks graded ``$\circ$'' are probable
detections but less prominent than those graded ``+'' and may be more
prone to projection effects. Candidates with a ``$-$'' grade are most
likely artifacts or maybe in some cases very poor clusters or groups.

Out of the total of $116$ matched-filter peaks $49$, were graded ``+'',
$42$ ``$\circ$'', and $25$ received a grade of ``$-$''. There is a
clear correlation between the significance of a cluster candidate and
its grade. The average SNR of a ``+'' rated cluster is $5.5\sigma$ and
include the highest SNR detections; ``$\circ$'' graded cluster
candidates have an average significance of $4.9$, while the unlikely
candidates rated with ``$-$'' have an average value of
$\overline{\sigma}_\mathrm{max} = 4.4$.

In the field RBS~0864, two clusters were detected independently on both
pointings on this field. These are \object{BLOX~J1023.6+0411.1}
(\object{RBS~0864} itself) and \object{BLOX~J1022.9+0411.9}. Their
properties reported in the table of matched-filter detected clusters
at the CDS are averages of both independent detections.

\begin{figure}[t]
  \resizebox{\hsize}{!}{\includegraphics{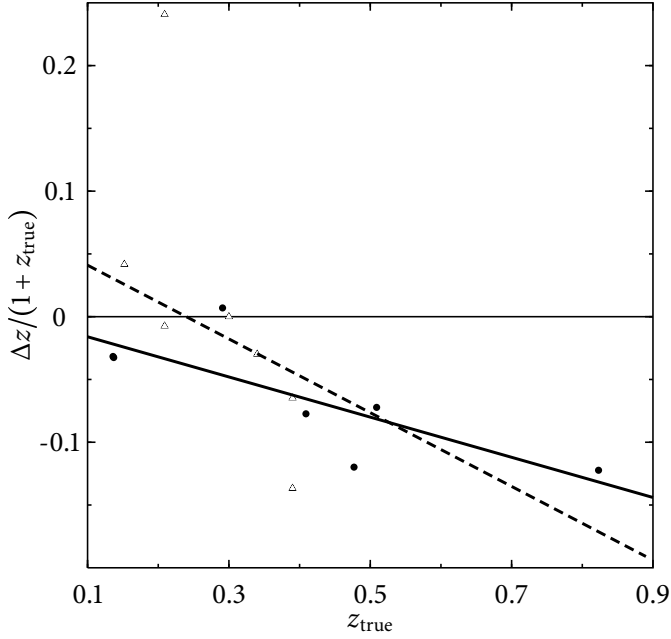}}
  \caption{Comparison of redshift estimates obtained from the
    \citetalias{1996AJ....111..615P} matched filter with literature
    values. Solid circles denote data points with spectroscopic
    redshift information; open triangles are photometric redshift
    estimates from the literature. The matched filter underestimates
    the redshift as is shown by the best-fit lines to the
    spectroscopic sample (solid line) and the full sample (dashed
    line).}
 \label{fig:match-zcomp}
\end{figure}

A comparison of the matched-filter estimated redshifts of the $13$
previously found systems with their redshift values found in the
literature is provided by Fig.~\ref{fig:match-zcomp}. Seven of these
$13$ clusters have spectroscopic redshift information, while the
remaining $6$ cluster have only photometric redshift estimates. We
point out that the redshifts of the photometric sample are not
photometric redshifts in the classical sense of, e.g., template
fitting but are the redshift estimators of other cluster finding
methods. This is most notably the redshift estimator of the
``cut-and-enhance method'' of \citet{2002AJ....123.1807G}, which
provides $4$ redshifts in this sample. The mean offset from the zero
line for the spectroscopic sample is marginally significant with
$\langle\Delta z/(1+z_\mathrm{true})\rangle = -0.06\pm0.05$, while the
mean offset of the whole sample $\langle\Delta
z/(1+z_\mathrm{true})\rangle = -0.06\pm0.07$ is consistent with zero,
where the error is simply the standard deviation of the scatter. The
relatively small deviation from zero hides a significant bias of the
matched filter redshifts towards lower redshifts. The solid line in
Fig.~\ref{fig:match-zcomp} is the best linear fit to the spectroscopic
data points. The line is described by $\Delta z/(1+z_\mathrm{true}) =
(0.00 \pm 0.03) + (-0.16 \pm 0.06) z_\mathrm{true}$. Thus, the
deviation from the ideal relation is significant at the $2.7\sigma$
level. The bias increases marginally if the photometric data points
are included in the analysis. The best-fit line is then described by
$\Delta z/(1+z_\mathrm{true}) = (0.07 \pm 0.05) +
(-0.29\pm0.12)z_\mathrm{true}$.

Since the clusters were found on different fields with independent
photometric calibration, errors in the photometric calibration can be
excluded as the source of this systematic difference. Possible sources
of the underestimated redshifts are (1) the $k$-correction, which
depends on the adopted galaxy model and (2) the luminosity function of
the fiducial cluster model, specifically the value of $M^*$. The
discrepancy increases with higher redshifts. Our flux filter was
constructed under the assumption of no evolution and thus stronger
discrepancies are indeed expected at higher redshifts. This has
already been noted by \citetalias{1996AJ....111..615P}, who also found
the estimated redshifts of higher redshift clusters to be
systematically too low by about $\Delta z = 0.1\ldots 0.2$.

\begin{figure}[t]
  \resizebox{\hsize}{!}{\includegraphics{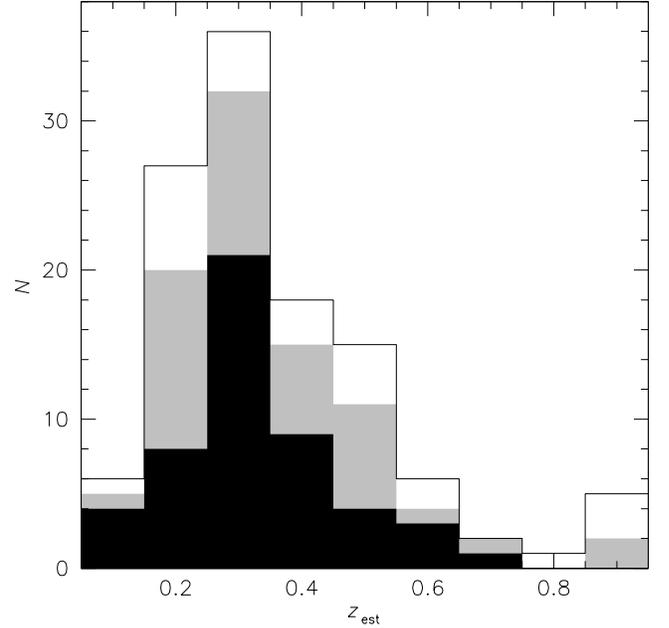}}
  \caption{Histogram of the estimated redshift distribution of the
    matched-filter detected clusters. The open histogram represents
    all cluster candidates, the gray histogram the ``+'' and
    ``$\circ$'' rated candidates, and the black histogram only the
    ``+'' rated clusters. }
\label{fig:match-hist}
\end{figure}

The \citetalias{2007A&A...461...81O} paper carried out a
matched-filter cluster search in the Canada-France-Hawaii Telescope
Legacy Survey (CFHTLS) Deep. The CFHTLS field D4 coincides with our
field LBQS2212$-$1759. We briefly compare the results of these two
independent cluster searches. They find 6 matched-filter selected
cluster candidates inside the WFI FOV. Two of their detections,
\object{CFHTLS-CL-J221500-175028} and
\object{CFHTLS-CL-J221537-174533}, coincide with matched-filter
cluster candidates, \object{BLOX~J2215.0$-$1750.5} and
\object{BLOX~J2215.6$-$1745.5}, detected in this survey. Both cluster
candidates received the best grade in either survey. The matched
filter redshift estimates agree for
\object{CFHTLS-CL-J221500-175028}/\object{BLOX~J2215.0$-$1750.5},
while the \citetalias{2007A&A...461...81O} put
\object{CFHTLS-CL-J221537-174533}/\object{BLOX~J2215.6$-$1745.5}
slightly higher at $z=0.4$ than our estimate of $z=0.3$. Two other
cluster candidates of \citetalias{2007A&A...461...81O} coincide with
possible clusters we found with other methods.
\object{CFHTLS-CL-J221620-173224}, an ``A'' rated potential cluster at
an estimated redshift of $z=0.7$ matches our weak lensing detection
(see Sect.~\ref{sec:weak-lens-detect}) \object{BLOX~J2216.3$-$1733.0}.
\object{CFHTLS-CL-J221606-175132}, a ``B'' rated cluster candidate
with a redshift estimate of $z=0.4$ matches our weak lensing detection
\object{BLOX J2216.1-1751.7}. Our matched-filter cluster candidate
\object{BLOX~J2214.4$-$1728.1} at $z=0.2$ is not found by
\citetalias{2007A&A...461...81O}.

\section{Weak lensing detection}
\label{sec:weak-lens-detect}
The tidal gravitational field of a galaxy cluster causes a coherent
tangential of the sheared images of background galaxies. A
quantitative measure for this alignment, the aperture mass $\map$, was
developed in \citet{1996MNRAS.283..837S}. There, $\map$ is defined as
a weighted integral over the dimensionless surface mass density
$\kappa$,
\begin{equation}
  \label{eq:1}
  \map(\vec{\theta}_0) = \int_{\mathrm{sup} U}
  \dif^2\theta\; U(\vartheta) \kappa(\vec{\theta})\;, 
\end{equation}
where $U(\vartheta) = U(|\vec{\theta}-\vec{\theta_0}|)$ is a radially
symmetric weight function with zero total weight to avoid the
mass-sheet degeneracy \citep{1995A&A...294..411S}. The aperture mass
in Eq.~(\ref{eq:1}) is defined in terms of the surface mass density,
but it is possible to find an expression that allows computing $\map$
in terms of the observable shear $\gamma$:
\begin{equation}
  \label{eq:2}
  \map(\vec{\theta}_0) = \int_{\mathrm{sup} Q} \dif^2\theta\;
  Q(\vartheta) \gamma_\mathrm{t}(\vec{\theta}; \vec{\theta}_0)\;,
\end{equation}
where we define the tangential shear $\gamma_\mathrm{t}$ relative to a
point $\vec{\theta}_0$ by
\begin{equation}
  \label{eq:3}
  \gamma_\mathrm{t}(\vec{\theta};\vec{\theta}_0) = -\Re\left[\gamma(\vec{\theta}
    +\vec{\theta}_0) 
    \ex^{-2\im\varphi}\right] \; .
\end{equation}
Here, $(\vartheta, \varphi)$ are polar coordinates with respect to
$\vec{\theta}_0$, and the weight function $Q(\vartheta)$ is related to
$U(\vartheta)$ by
\begin{equation}
  \label{eq:4}
  Q(\vartheta) = \frac{2}{\vartheta^2} \int_0^\vartheta
  \dif\vartheta^\prime\; \vartheta^\prime U(\vartheta^\prime) -
  U(\vartheta) \; .
\end{equation}
On real data the aperture mass can be estimated by a sum over the
$N_\mathrm{g}$ galaxy ellipticities inside the aperture,
\begin{equation}
  \label{eq:5}
  M_\mathrm{ap}(\vec{\theta}_0) = \frac{1}{n} \sum_{i=1}^{N_\mathrm{g}}
  Q(\vartheta_i) \varepsilon_{\mathrm{t}i} \; .
\end{equation}
Here $n$ is the number density of faint background galaxies (FBG), and
the tangential component of the ellipticity has been defined in
analogy to Eq.~(\ref{eq:3}).

The SNR of a peak in maps of aperture mass can be estimated by using
the fact that $\langle\map\rangle \equiv 0$ holds in the case of no
lensing. Then the RMS dispersion of the $\map$-statistic becomes
$\sigma_{\map} = \sqrt{\langle \map \rangle^2}$, which is
\begin{equation}
  \label{eq:6}
  \sigma_{\map} = \frac{\sigma_\varepsilon}{\sqrt{2}n} \left[
    \sum_{i=1}^{N_\mathrm{g}} Q^2(\theta_i)\right]^{1/2} \; ,
\end{equation}
using
\begin{equation}
  \label{eq:7}
  \langle \varepsilon_i \varepsilon_j \rangle =
  \frac{\sigma_\varepsilon^2}{2}\delta_{ij} \; .
\end{equation}

What remains to be fixed is the shape of the weight function. If only
white noise caused by the random ellipticity of background galaxies is
present, i.e., if the noise is described by Eq.~(\ref{eq:6}), the
filter function $Q$ should follow the shear profile of the cluster as
closely as possible to increase the SNR of a galaxy cluster detection.
The aperture mass then becomes a \emph{matched filter technique} for
weak lensing detections of galaxy clusters. 

$N$-body simulations predict the shape of collapsed halos to follow an
NFW profile. The shear of an NFW profile can be computed analytically
\citep{1996A&A...313..697B,2000ApJ...534...34W}, but the resulting
expressions are complex and time-consuming to evaluate.
\citet{2007A&A...462..875S} have proposed a simple approximation to the
NFW shear profile,
\begin{equation}
  \label{eq:8}
  Q_{\mathrm{NFW}}(x) = \frac{\tanh x}{x}\;,
\end{equation}
that is much faster to compute than the full expression. The foregoing
equation is not used directly for the computation of $\map$, but
exponential cut-offs are multiplied to Eq.~(\ref{eq:8}) as
$x\rightarrow0$ and $x\rightarrow\infty$,
\begin{equation}
  \label{eq:9}
  Q_{\mathrm{NFW}}(x;x_\mathrm{c}) =
  \frac{1}{1+\ex^{6-150x}+\ex^{-47+50x}} \frac{\tanh
    (x/x_\mathrm{c})}{x/x_\mathrm{c}}\;.
\end{equation}
The purpose of these cut-offs is (1) to avoid finite field effects as
the weight function~(\ref{eq:8}) formally extends to infinity, but
data is only available on a finite field; and (2) to downweight the
signal close to the cluster core where cluster dwarf galaxies may
pollute the signal and where the reduced shear $g$ rather than the
shear $\gamma$ should be used. The parameter $x_\mathrm{c}$ in
Eq.~(\ref{eq:8}) controls the shape of the filter function. For low
values of $x_\mathrm{c}$, more weight is put to smaller filter scales.
In the absence of the exponential cut-offs, variations in
$x_\mathrm{c}$ and $\theta_\mathrm{max}$ are obviously degenerate, but
the exponential cut-off makes the weight function~(\ref{eq:9}) a
two-parameter family. \citet{2005A&A...442...43H} find that the
optimal value of $x_\mathrm{c} = 0.15$ for a range of cluster masses
and redshifts, so we will use this value throughout this work.

The projection of large-scale structure (LSS) along the line of sight
can potentially be a serious contaminant for every weak lensing
observation of galaxy clusters. Such projections of sheets and
filaments inevitably lead to false cluster detections at all
significance levels expected from real clusters
\citepalias{2005ApJ...624...59H} and missed cluster detection except
for the most massive systems \citepalias{2004MNRAS.350..893H}. It was
shown by \citetalias{2005ApJ...624...59H} that, even in the absence of
shape noise, the efficiency of a weak lensing search for galaxy
clusters does not exceed $\sim 85\%$ and also depends on the shape of
the filter function.

\citet{2005A&A...442..851M} propose to treat the projections of
masses in the background and foreground of the cluster, i.e., the
large-scale structure as a source of non-white noise. They constructe
an optimal filter that tries to maximize the signal caused by a galaxy
cluster while downweighting the cosmic shear signal on scales of
interest. The resulting filter function $Q_\text{LSS}$ depends not
only on the expected shear profile of the galaxy cluster but also on
the number density of FBG and on the convergence power spectrum to
describe the expected noise. The latter noise contribution in turn
depends on the redshift distribution of the FBG. In the case of pure
white noise, the Maturi filter takes the shape of an NFW shear
profile, i.e., essentially the same form as the Schirmer filter, and
both filters are virtually equivalent. If projections of the LSS are a
significant noise source, then using $Q_\text{LSS}$ instead of
$Q_\text{NFW}$ can lead to significantly lower contaminations of the
cluster catalog with false positives.

\subsection{Signal and noise of the aperture mass}
\label{sec:sign-noise-apert}
Before we apply the aperture mass statistic to the XFS data we need to
understand the properties of the aperture mass function in some more
detail. Equations~(\ref{eq:5}) and~(\ref{eq:6}) allow us to compute
the SNR of $\map$ as
\begin{equation}
  \label{eq:10}
  S = \frac{\sqrt{2} \sum_i Q_i
    \varepsilon_{\mathrm{t}i}}{\sqrt{\sum_i Q_i^2
      \varepsilon_i^2 }}\;, 
\end{equation}
where $Q_i$ is the weight assigned to the $i$th galaxies by the
radially symmetric weight function $Q(\vartheta)$. This expression,
however, considers only the noise caused by the random ellipticities
and not the shot noise of the finite sampling of the $\map$ statistic
by the population of background galaxies. 

We used ray-tracing simulations to examine these effects in detail.
These were generated by tracing light rays through the
$\Lambda$CDM $N$-body simulation of the VIRGO consortium
\citep{1998ApJ...499...20J}. These $N$-body simulations have been
carried out with the following parameters: $\Omega_\mathrm{m} = 0.3$,
$\Omega_\Lambda = 0.7$, $\h = 1$, $\sigma_8 = 0.9$, $\Gamma = 0.21$,
and the index of the primordial power spectrum $n_\mathrm{s} = 1$. The
population of background galaxies has been assumed to be a
$\delta$-function peaked at $z = 1$.

We now give a short description of our ray-tracing algorithm. More
details of can be found in \citet{2005Dipl...1....1H}. From redshift
$0$ to the source redshift, $1024^2$ light rays are traced through
$16$ slices of $202.9\,\hm$\,Mpc thickness onto an output grid of
$1\times 1$\,square degree. Each redshift slice corresponds to one
output box of the $N$-body simulation and is projected as a whole onto
a lens plane, preserving the periodic boundary conditions of the
$N$-body box. To avoid repetition of structure along the line of
sight, the planes are randomly shifted and rotated. The light rays are
shot from the observer through the set of lens planes, forming a
regular grid on the first plane. We then use FFT methods to compute
the lensing potential on each lens plane, from which we obtain the
deflection angle and its partial derivatives on a grid. The ray
position and the Jacobian of the lens mapping for each ray are
obtained by recursion. Given the ray position on the current lens
plane, its propagation direction (known from the position on the last
plane), and the deflection by the current plane interpolated onto the
ray, we immediately obtain the ray position on the next plane.
Differentiation of this recursion formula with respect to the image
plane coordinates yields a similar relation for the Jacobian of the
lens mapping, taking the tidal deflection field computed before into
account. The recursion is performed until we reach the source plane.
From the final Jacobian, we obtain noiseless convergence and shear
maps in the usual way.

Since we wish to study which convergence/$\map$ peaks correspond to
real dark matter halos, we also have to compute the lensed positions
of the central particles of the dark matter halos contained in the
halo catalogs. We achieve this by projecting the halo position onto
the lens planes and identifying the light ray that passes closest to
the halo. The lensed position of the halo is then computed by
inverting the linearized lens mapping around this ray. 

Fifty different realizations were made by using in each case different
random shifts and rotations of the lens planes. Lensing catalogs were
created by randomly distributing galaxies with an ellipticity
dispersion of $\sigma_\varepsilon = 0.38$ over the output grid of the
ray-tracing simulations on areas corresponding to the sizes of actual
XFS fields until the number density of the respective XFS field was
reached. When placing galaxies, the masks used in the catalog
generation of the real data were applied to simulated catalogs as
well, to realistically model the effect of holes in the field. Three
different masks and number densities were taken from the XFS data. All
masks were put on each of the $50$ ray-tracing realization to obtain a
high number of lensing simulations. The effective (unmasked) area and
number densities of these fields are $970$\,arcmin$^{2}$ and
$13.4$\,arcmin$^{-2}$, $1022$\,arcmin$^{2}$ and $17.7$\,arcmin$^{-2}$,
and $981$\,arcmin$^2$ and $18.2$\,arcmin$^{-2}$. The total area
covered by our ray-tracing simulations is $41.3$\,sq.\,deg.

With typical sizes of the XFS fields of $35\arcmin\times 35\arcmin$,
the individual lensing simulations are not totally independent but
have some overlap because the side length of one ray-tracing simulation
is only $60\arcmin$, not enough to accommodate two XFS masks next to
each other. The catalogs were placed on the ray-tracing simulations
such that this overlap is minimized. While not completely independent,
these overlapping areas were covered by catalogs with different
maskings, different number densities, and different realizations of
Gaussian noise so that for our purpose -- understanding the noise
properties of $\map$ from realistic simulations -- no significant
correlation between individual lensing simulations is expected.

\begin{table}[t]
 \centering
 \caption{Filter radii for $\map$ computation and corresponding
  virial mass.}
 \begin{tabular}{rr@{.}l}
  \hline\hline
  Radius & \multicolumn{2}{c}{$M_\mathrm{vir}$} \\
  (kpc/$\h$) & \multicolumn{2}{c}{($10^{14}\,\hm\,M_{\sun}$)} \\
  \hline
  1000 & 0&76 \\
  1247 & 1&5 \\
  1493 & 2&5 \\
  1740 & 4&0 \\
  1986 & 6&0 \\
  2233 & 8&5 \\
  2479 & 12&0 \\
  2726 & 15&0 \\
  2972 & 20&0 \\
  \hline
 \end{tabular}
 \label{tab:filter_radii}
\end{table}

We computed the aperture mass for $9$ different filter scales and the
filter functions proposed by \citet{2007A&A...462..875S} and
\citet{2005A&A...442..851M} from the same catalogs. As both filter
functions are based on an NFW model of a cluster, we chose the filter
radii based on an assumed fiducial cluster model. In this model the
cluster is at redshift $z=0.3$, the redshift at which we expect the
lens strength in our survey to be maximized.
Table~\ref{tab:filter_radii} gives these filter radii and the
corresponding virial mass if the filter scale is interpreted as the
virial radius of the cluster. We also need to fix the redshift of the
source galaxies to model the large-scale structure power spectrum in
the Maturi filter. We assumed that all background galaxies are at
$z_\mathrm{s} = 0.8$ and that the number density of background
galaxies is $n = 18$\,arcmin$^{-2}$. We computed $\map$ on a grid and
set the pixel size in this grid such that one pixel corresponds to
$50\,\hm$\,kpc at the redshift of the fiducial cluster model.

For all lensing simulations and filter scales we computed $\map$ and
$-\map$ (for later detection of negative peaks as a control) for both
filter functions from the input catalog and $\map$ after rotating (1)
all galaxies by $45^\circ$; (2) every galaxy by a random angle for the
Schirmer filter. Maps of the aperture mass and their SNR
(\emph{$S$-maps}), which were computed as well, were saved as FITS
images.

As for the detection of peaks in the matched filter maps, we used
\texttt{SEx\-trac\-tor} to identify shear-selected peaks in the
$\map$-maps. For this purpose we run \texttt{SEx\-trac\-tor} in
dual-image mode with the $\map$-map as the detection image and the
$S$-map as the measurement image. This means that the SNR of a $\map$
peak is computed with Eq.~(\ref{eq:10}) and not determined by
\texttt{SExtractor}. This is a small difference from the matched
filter pipeline, in which we did not use the
\citetalias{1996AJ....111..615P} likelihood to determine a
significance but used \texttt{SExtractor} detection significances.

The detection threshold is set to $2\sigma$ and the minimum detection
area scales with the filter such that it corresponds to the pixels
covered by a circle with a radius twice as large as the filter scale.
\texttt{SExtractor} is run without deblending, i.e., every contiguous
area above the detection threshold is counted as one object. Peaks
found in this way in different filter radii were associated based on
positional coincidence.

We first examine the results obtained with the Schirmer filter before
comparing these to the ones obtained with the Maturi filter.
Positional offsets between the weak lensing peak positions and the
position of the BCG or the center of X-ray emission from the hot
intracluster gas are commonly observed \citep{2006ApJ...643..128W} and
expected \citep{2005A&A...440..453D}. The size of the offsets has been
studied for the case of an isolated SIS by
\citet{2005A&A...440..453D}. The ray-tracing simulations used here
allow us to investigate the additional effect of large-scale structure
along the line-of-sight.

The catalogs of weak lensing halos produced in this way from the
$\map$ images were associated with a catalog of actual dark matter
halos in the VIRGO simulation. For this purpose we only considered
halos with masses in excess of $10^{14}\,M_{\sun}$ and with redshifts
$0.1 < z < 0.7$ as these are roughly the ones to which we expect
to be sensitive in our galaxy cluster survey. The maximum distance
allowed for a match between halo position and $\map$ peak was
determined by the virial radius of the halo and the size of the $\map$
peak as determined by \texttt{SExtractor}. Note that the position of
the halo was derived from the most-bound particle in that halo
identified by the friend-of-friend halo finder employed to generate
the halo catalog. In rare cases this is in the center between what 
by eye be would identified as two separate halos.

\begin{figure}[t]
  \resizebox{\hsize}{!}{\includegraphics{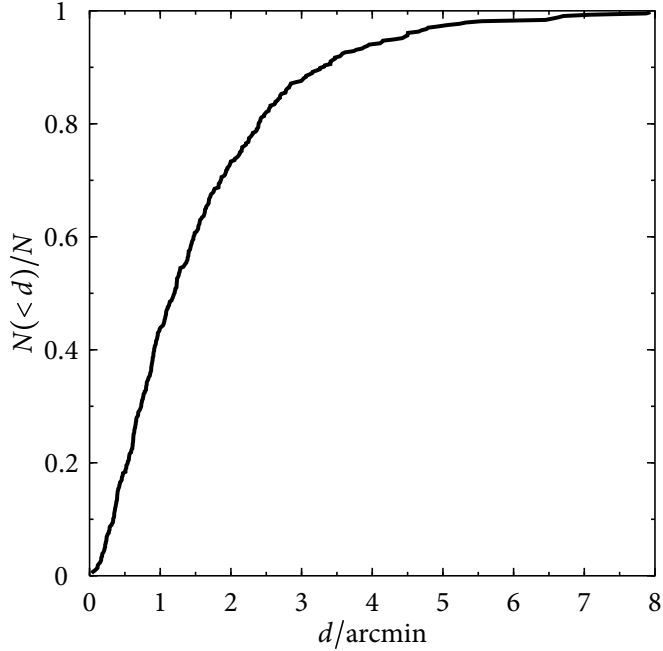}}
  \caption{Cumulative distance distribution of the $434$ $\map$ peaks
    that could be associated to a dark matter halo in the VIRGO
    simulation. $75\%$ of all matches are made within a $2\farcm15$
    radius.}
 \label{fig:cumul-dist_rayt}
\end{figure}

In the ray-tracing simulations, $434$ peaks in the $\map$-maps could be
associated with dark matter halos in the VIRGO simulation.
Figure~\ref{fig:cumul-dist_rayt} shows the cumulative distribution of
their positional differences. From the number density of $\map$
peaks and the average size of the association radius we estimate that
$\sim 100$ or $\sim 25\%$ of those matches are chance coincidences.
Inspection of Fig.~\ref{fig:cumul-dist_rayt} shows that $75\%$ of all
positional offsets are smaller than $2\farcm15$, which is the maximum
offset we will allow from here on. Note that on the one hand this is 
significantly smaller than the $3\arcmin$ matching radius adopted by
\citetalias{2005ApJ...624...59H}, who used a higher number density and
a smaller ellipticity dispersion. On the other hand, it is
significantly larger than the offsets found for an isolated SIS by
\citet{2005A&A...440..453D}. This possibly indicates a non-negligible
influence of large-scale structure along the line of sight on the weak
lensing peak positions of dark matter halos. One, however, has to be
careful when drawing this conclusion as we are looking at halos at
very different redshifts, while \citet{2005A&A...440..453D} studied
only systems at one redshift.

Aperture mass peaks not associated with dark matter halos can be
caused either by projections of large-scale structure mimicking a
shear signal of a cluster or by the shape noise of background galaxies
that can cause random tangential alignments. In real data the measured
ellipticities must be corrected for atmospheric seeing and PSF
distortions of the instrument. Residuals in this correction can lead
to spurious alignments of background galaxies, both in curl-free shear
fields (pure \emph{E-modes}) and in non-curl free shear fields (also
including \emph{B-modes}). E-modes are transformed into B-modes (and
vice versa) by rotating all galaxies by $45^\circ$. Because
gravitational lensing only creates E-modes, the observed power of
B-modes is often used as a quality check of the PSF correction. We did
not check our lensing signal for the presence of B-modes. However,
cosmic shear studies using PSF correction schemes very similar to ours
find that B-modes typically occur on scales smaller than are relevant
for our $\map$ kernels
\citep[e.g.,][]{2005A&A...429...75V,2006astro.ph..6571H}. It is thus
safe to assume that we can apply the results of our ray-tracing
simulations to the XFS.

\begin{figure}[t]
  \resizebox{\hsize}{!}{\includegraphics{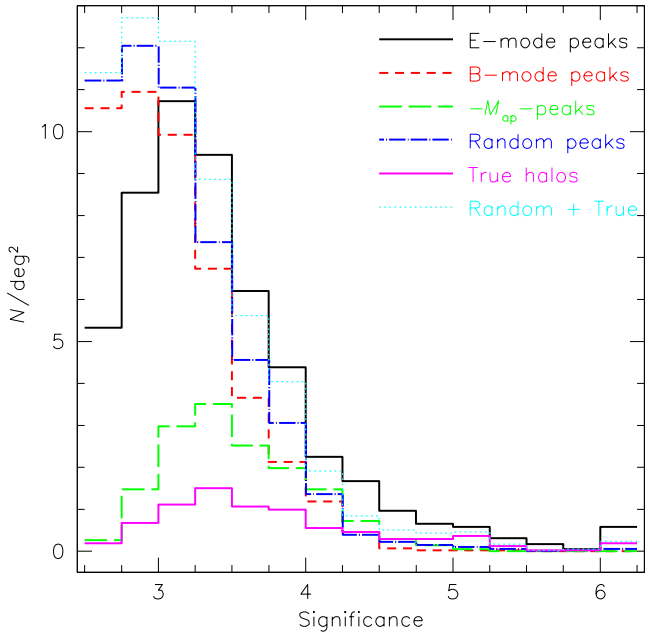}}
  \caption{Peak significances of shear selected peaks in the simulated
    maps of $\map$ (black solid), $-\map$ (green long dashed), B-mode
    $\map$ (red short dashed), and mock catalogs (blue dot-dashed).
    The solid pink line corresponds to peaks successfully associated
    with dark matter halos within $2\farcm15$. The thin light-blue
    dotted line is the sum of the random peaks (blue) and the true
    halos (pink).}
\label{fig:rayt_significances}
\end{figure}

Figure~\ref{fig:rayt_significances} shows the significance
distribution of shear-selected peaks in the different kinds of
aperture mass maps created with the Schirmer filter. These are (1) the
distribution of peak significances for all $\map$ peaks created from
the ray-traced catalog, i.e., those peaks one would find in real data,
(2) negative peaks or, considering how the peak finding pipeline is
run, peaks found in $-\map$-maps, (3) peaks found in B-mode
$\map$-maps, i.e., maps of aperture mass created after all galaxies in
the ray-traced catalog were rotated by $45^\circ$, (4) weak lensing
peaks that could be associated with dark matter halos within a
matching radius of $2\farcm15$, and (5) peaks found in mock catalogs
with random ellipticities. It is important to emphasize that the
random seed was kept fixed so that when computing $\map$ on different
filter scales the orientation of galaxies in the input catalog
remained unchanged. Here we consider only those peaks that were
detected on at least two filter scales. We justify this choice later
when examining the influence of the number of filter scales in which
peaks are detected. The significance used in the histogram is the
maximal significance values of all filter scales a peak was detected
in.

A number of features in Fig.~\ref{fig:rayt_significances} are worth a
more detailed discussion. First, we note that the number density of
peaks in the $\map$- and $-\map$-maps declines towards lower
significances, while in the B-mode and random maps it remains roughly
constant below $3.25\sigma$. Naively, one would expect an increase in
all curves towards lower significances. The observed behavior is due
to a selection bias when running \texttt{SExtractor} on the FITS
images. The detection threshold is derived from the standard deviation
of the background in these images. The B-mode and random maps are
overall flatter than the E-mode $\map$-maps. This leads to a detection
threshold at which peaks of lower significance in the S-maps are
detected in the $\map$-maps for random maps and B-mode maps than for
E-mode maps.

Second, the number densities of random and B-mode
peaks are very similar, with the latter slightly lower. As we
did not simulate the systematic influence of instrumental PSF
corrections, the only sources of B-modes are the shape noise of
background galaxies and finite field effects in the $\map$ estimator.
The effect of shape noise alone is simulated by the random catalogs,
while the B-mode peaks are a combination of shape noise and systematic
effects due to the finiteness of the field and holes in the data. The
fact that the number density of B-mode peaks is compatible with the
number density of random peaks shows that systematic effects
contributing to B-modes are not an important noise source. This is
confirmed by a visual inspection of the peak distribution indicating
that B-mode peaks do not show an obvious tendency to appear close to
holes in or edges of the data field.

Third, the number of E-mode peaks is higher than the number of any
other peak statistic in all significance bins $>3.25\sigma$ in which
the aforementioned selection plays no role. The sum of the true halos
peaks and the random peaks is compatible with the number of E-mode
peaks in the significance bins from $3.25\sigma$ to $4.25\sigma$, if
one assumes Poissonian statistics. At higher significances, an excess
of E-mode peaks is observed. We surmise that this is due to
projections of large-scale structures. This means that we can expect a
significant fraction of spurious peaks at almost all significances, a
result that is compatible with earlier findings of
\citetalias{2004MNRAS.350..893H} and \citetalias{2005ApJ...624...59H}.
At low significances the spurious peaks will be dominated by shape
noise, while at high significances many spurious peaks will be caused
by the projection of large-scale structures. We emphasize that the
latter class of peaks is in fact caused by gravitational lensing. They
just do not correspond to a single mass concentration in 3-d space.
These peaks are spurious peaks only in the sense of a galaxy cluster
search.

Fourth, negative peaks are relatively rare. The two effects leading to
this result are best understood in terms of the filter function $U$
that is related to Schirmer's $Q_\mathrm{NFW}$ function by
Eq.~(\ref{eq:4}) and acts on the surface mass density. The function
$U$ has a narrow positive peak close to the origin with extended and
shallow negative wings to satisfy the condition that $U$ has zero
total weight. The comparably low number density of $-\map$ peaks is
then caused by (1) the shallowness of negative wings, which will limit
the peak strengths of negative peaks; and (2) by $U$ acting as a
bandpass filter for structures with the same size as the
characteristic filter scale. The large extent of the negative wings
will make negative peaks more extended than positive peaks, naturally
leaving less space for other peaks.

\begin{figure}[t]
  \resizebox{\hsize}{!}{\includegraphics{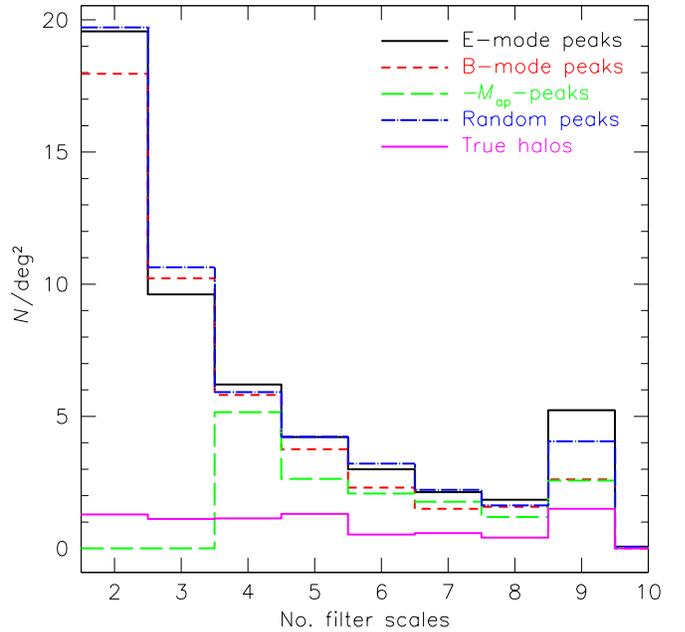}}
  \caption{Number density of lensing peaks in dependence on the number
    of filter scales in which the peak is detected. The
    color/line-style coding is the same as in
    Fig.~\ref{fig:rayt_significances}. Dark matter halo peaks show
    virtually no dependence on the number of filter scales, while the
    number density of spurious peaks sharply declines with a
    requirement on the minimum number of filter scales.}
\label{fig:rayt_slices}
\end{figure}

By associating peaks found in different filter scales with each other,
we can also examine whether the number of filter scales $n_\mathrm{f}$
that a peak is detected in says something about the correspondence of
the lensing signal to a dark matter halo. Figure~\ref{fig:rayt_slices}
shows the number density of peaks detected in the ray-tracing
simulations as a function of $n_\mathrm{f}$. The color/line-style
coding is the same as in Fig.~\ref{fig:rayt_significances}. One
clearly sees that the number density of lensing peaks associated with
a dark matter halo is virtually independent of the number of filter
scales the $\map$ peak appears in. Real dark matter halos show up as
often in only $3$ filter scales as they do in $9$ filter scales.

The behavior of spurious peaks is very different. Their number falls
off monotonically as a function of $n_\mathrm{f}$, with an exception
in the last bin, i.e., the peaks that are detected in all filter
scales. This dependence can be used to impose a selection criterion to
decrease the contamination of shear selected clusters with spurious
lensing peaks by requiring that a $\map$ peak must occur in at least a
given number of filter scales. However, as stated above, this will
always exclude a number of real clusters as well. As a compromise
between efficiency and completeness, we imposed the condition that
$n_\mathrm{f} \ge 3$ in the XFS.

We now briefly compare these results to the ones obtained with the
Maturi filter on the same catalogs. This filter can potentially result
in fewer spurious peaks and increase the SNR of real clusters, thereby
increasing the number of real cluster detections. We find that the
results of the Schirmer and Maturi filters are entirely consistent for
our survey parameters. The number of $\map$ peaks associated to dark
matter halos within the $2\farcm15$ matching radius ($316$) is,
contrary to expectations, slightly smaller than for the Schirmer
filter ($325$). The difference might well be due to a lower
contamination with spurious peaks at high significances where the
contamination with LSS projections should be suppressed by the Maturi
filter. The observed peak offsets from the dark matter halo positions
are consistent with the those of the Schirmer filter. We compared the
peak significances of $\map$ peaks related to dark matter halos
computed with both filters and find $\langle S(Q_\mathrm{LSS}) -
S(Q_\mathrm{NFW})\rangle = -0.06 \pm 0.27$, i.e., the peak
significances are consistent with each other, non-significantly
favoring the Schirmer filter. Again, this small difference might be
due to the relative suppression of highly significant projections of
the LSS. These findings agree with \citet{2007A&A...462..473M} who
compare the performance of their filter with the cluster search by
\citet{2007A&A...462..875S} on the same fields. As a result of our
comparison we limit the weak lensing cluster search in the XFS to the
Schirmer filter that is less complex and faster to compute.

\subsection{Weak lensing catalog}
\label{sec:weak-lensing-catalog}
Based on the optical catalogs created with \texttt{SExtractor}'s
default convolution kernel (Sect.~\ref{sec:optic-catal-creat}), we
created lensing catalogs. We used the KSB algorithm
\citep{1995ApJ...449..460K} to obtain shear estimates closely
following the procedure described by \citet{2001A&A...366..717E}. From
the KSB catalogs we constructed catalogs of probable background galaxies
that are reliable shear estimators by imposing the following selection
criteria: Objects with SNR$<2$, Gaussian radius $r_\mathrm{g} <
0\farcs33$ or $r_\mathrm{g}> 1\farcs19$, or corrected ellipticity
$\varepsilon > 1.0$ were deleted from the catalog. We also deleted
objects whose pre-seeing shear polarizability tensor $P^\gamma $ has a
negative trace, and bright galaxies with $R<21$\,mag.

Table~\ref{tab:optical_fields} gives the effective (unmasked) area of
all XFS fields used for the weak lensing cluster search, as well as
their number density of galaxies in the weak lensing catalogs. The
total area used for weak lensing is $6.4$\,sq.\,deg. The number
density of background galaxies averaged over this area is
$14.1$\,arcmin$^{-2}$.

The aperture mass statistic with the Schirmer filter function was
estimated from the resulting catalogs on the filter scales listed in
Table~\ref{tab:filter_radii}. No weighting of individual galaxies was
done. Peaks in the $\map$-maps were detected as described in the
previous section. The final catalog lists all aperture mass peaks with
maximum SNR $\ge 5$, or with maximum SNR $\ge 3$ if the peak has an
X-ray or matched-filter counterpart, or was previously reported as
cluster (candidate) in the literature within $2\farcm15$
radius from the lensing position. All shear peaks must be present in
at least $3$ filter scales.

In the following we give comments on some specific fields.

\begin{itemize}
\item \object{T Leo} -- No association could be found for the $4$
  $\map$ peaks with $\sigma_\mathrm{max}>3$ and $n_\mathrm{f}>2$ in
  this field. Thus, none of these peaks are included in our catalog.
\item Field~864-1 -- None of the $5$ $\map$ peaks with
  $\sigma_\mathrm{max}>3$ and $n_\mathrm{f}>2$ could be associated
  with X-ray or matched-filter clusters. These peaks are not included
  in our catalog.
\item Field~864-9 -- This field contains one of the three weak-lensing
  selected cluster candidates that was previously identified in the
  literature as a cluster candidate but not found by the matched
  filter or in X-ray. The candidate \object{BLOX J1343.5$-$0022.8} is
  outside the FOV of XMM-Newton. The cluster candidate is very
  elongated and might for this reason be missed by the matched filter
  pipeline.
\item \object{LBQS~2212$-$1759} -- This field contains the remaining
  two weak-lensing cluster candidates that match only clusters
  previously reported in the literature, but were found with neither
  the matched filter nor the X-ray survey. Both cluster candidates
  match optically selected clusters from \citetalias[][see also
  Sect.~\ref{sec:postm-match-filt}]{2007A&A...461...81O}.
  \citet{2007A&A...462..459G} have performed a weak lensing cluster
  search on this field as part of the CFHTLS Deep fields and have not
  found any convergence peak in the FOV of WFI above their detection
  threshold of $3.5\sigma$.
\end{itemize}

We selected a total of $31$ cluster candidates using the aperture mass
method on $23$ WFI fields. The full catalog of shear selected cluster
candidates is available in electronic format at the CDS. There we list
the position of the $\map$ peaks, the significance of their detection,
the number of filter scales a cluster was detected in, and the filter
scale in which the SNR of a peak was maximized. This catalog contains
4 of the 5 clusters that were primary XMM-Newton targets in the XFS
fields. The fifth cluster, \object{MKW 9}, is at redshift $z=0.04$,
too low to be detectable with weak lensing.

On average we find $1.3$ weak lensing cluster candidates per XFS
field. Most $\map$ peaks do not correspond to a cluster candidate.
This is to be expected from the results of our ray-tracing simulations
and illustrated by Figs.~\ref{fig:rayt_significances}
and~~\ref{fig:rayt_slices}. The two fields containing only unmatched
$\map$ peaks are statistically expected. The absence of weak lensing
peaks in these fields does not hint at problems with their lensing
catalogs.

\section{Summary and discussion}
\label{sec:summary}
In this work we have selected galaxy cluster candidates
independently with three different methods: optical matched filter
algorithm, extended X-ray emission, and the shear signal induced by
massive foreground structures.

We found a total of $155$ cluster candidates in $23$ WFI fields, or
$24.2$ cluster candidates per square degree. Most cluster candidates
were found with the optical matched filter ($116$), followed by X-ray
emission ($59$). As was previously shown
\citepalias{2004MNRAS.350..893H,2005ApJ...624...59H} and confirmed by
our lensing simulations using ray-tracing simulations
(Sect.~\ref{sec:sign-noise-apert}), the efficiency of weak lensing for
cluster selection is relatively low. To avoid being dominated by
spurious weak lensing signals, we limited the catalog of weak lensing
selected cluster candidates to those that have an optical or X-ray
counterpart, either found in our own survey or previously reported in
the literature. We found significant lensing signals for $31$ cluster
candidates, of which $12$ are previously known cluster (candidates).
Eleven of the weak lensing selected clusters were detected with both the
matched filter and X-ray emission, excluding \object{A~1882} that is
not part of our X-ray catalog; $6$ of these are previously unknown
cluster candidates.

Comparing the redshift estimates of the Postman matched-filter method
to spectroscopically measured redshifts or other photometric
estimates, we find that these work surprisingly well. The mean
difference in redshifts $\langle \Delta z/(1+z_\mathrm{true})\rangle$
is marginally significant only for the spectroscopic sample with
$-0.06\pm 0.05$, and consistent with zero if we also trust the
redshifts of the photometric sample that gives $\langle \Delta
z/(1+z_\mathrm{true})\rangle = 0.06 \pm 0.07$. Considering that we use
only one passband to derive the redshift of matched-filter clusters,
this result compares favorably to what is achievable with more colors.
For example, \citet{2002AJ....123.1807G} report a mean $\Delta z$ for
their ``cut-and-enhance'' method of $0.02$ using four colors, but only
after outliers with $\Delta z > 0.1$ have been rejected. However, the
matched filter estimated redshifts come with a significant bias, which
puts higher redshift cluster at redshifts that are typically too low
by $\Delta z = 0.1\ldots0.2$.

We described in detail how we developed the selection criteria for our
weak lensing sample using ray-tracing simulations. We find that -- at
least for the comparatively low number densities and peak
significances we are dealing with -- the dominant source of noise is
the shape noise of the background population and not projections of
the large-scale structure. Our ray-tracing simulations suggest that
the contamination with projections of the large-scale structure
becomes more important at higher significances. However, the area
covered by the XFS is comparatively small and the absolute number of
highly significant $\map$ peaks is consequently small. The $\map$
kernel developed by \citet{2005A&A...442..851M} to minimize the effect
of large-scale structure on weak lensing cluster searched thus could
not perform better than the filter function proposed by
\citet{2007A&A...462..875S} used in this work. It is sensible to
assume that the Maturi filter will perform better on deeper surveys
\citep{2007astro.ph..2031P} for two reasons: (1) The higher number
density of FBG will reduce the white noise component and (2) a deeper
survey will probe a larger volume and hence increase the contribution
of LSS projections.

The matching radius of $2\farcm15$ we deduced from our ray-tracing
simulations is considerably smaller than the $3\arcmin$ employed by
\citetalias{2005ApJ...624...59H}. Considering that
\citetalias{2005ApJ...624...59H} used a higher number density of FBG
with an ellipticity dispersion lower by a factor of $1.6$, even their
already low efficiency estimates for weak lensing cluster searches
still seem to be too optimistic.

The number density of weak-lensing selected cluster candidates is
$4.8$ per square degree. This is slightly lower than the number
density of $\map$ peaks with a halo counterpart in the ray-tracing
simulations, which is $6.1/$sq.\,deg. However, the average number
density of background galaxies in the XFS data is only
$14.1$\,arcmin$^{-2}$; this is somewhat lower than in the ray-tracing
simulations, which had an average number density of
$16.5$\,arcmin$^{-2}$. Whether the difference in cluster counts can
really be attributed to the difference in number density should be
checked by adjusting the simulation parameters to match the XFS
observations better. The trend to slightly lower number densities is
also present if we select only $\map$ peaks with a higher SNR$\geq 4$.
The XFS contains $11$ of these highly significant peaks associated
with a matched filter or X-ray cluster candidate. This corresponds to
$1.7/$sq.\,deg., compared to $2.3/$sq.\,deg. in the ray-tracing
simulations. These high significance peaks include all 4 clusters that
were primary XMM-Newton targets at redshifts accessible by weak
lensing. These 4 clusters alone contribute a number density of
$0.6/$sq.\,deg.

With the cluster sample presented here we have built a solid
foundation for studying possible selection effects in either method.
This will be the subject of a follow-up paper (Dietrich et al., in
preparation).

\acknowledgement{We thank the anonymous referee for many comments that
  helped to improve the clarity of this paper. This work was supported
  by the German Ministry for Science and Education (BMBF) through DESY
  under the project 05AE2PDA/8, by the Deutsche Forschungsgemeinschaft
  under the project SCHN 342/3--1, by the German DLR under contract
  number 50OX0201, and by the Priority Program 1177 of the Deutsche
  Forschungsgemeinschaft}

\bibliographystyle{aa}
\bibliography{BoLOX-I}

\appendix

\section{X-ray observations}
\label{sec:x-ray-observations}
Table~\ref{tab:xray-observations} gives a summary of the EPIC X-ray
observations contributing to the XFS data used in this work. The table
gives for each field: in Col.~1 the field name; in Col.~2 the
XMM-Newton observation ID; in Col.~3 the nominal exposure time in
seconds; in Col~4--6 the settings for each of the cameras. Here (E)FF
indicates (extended) full frame readout, LW large-window mode, and SW
small-window mode. These cameras and their settings are described in
detail in \citet{2004.xmm.guide.E}. For some fields additional
observations were available, but these were discarded mainly due to
unsuitable camera settings.

``Field 864'' is a mosaic of $3\times 3$ XMM-Newton observations. The
name was assigned by the principal investigator. Fields~864-2, 4, 5,
6 are listed in Table~\ref{tab:xray-observations} although they are
not covered by our WFI observations. The reason is that these data
overlap with the Fields~864-6, and 9, which are XFS survey fields.

\begin{table*}
  \caption{Information about X-ray imaging in the XFS.}
  \begin{tabular}{ll>{$}r<{$}rrr}
    \hline\hline
    Field & Obs. ID & T_\mathrm{exp}/\mathrm{s} &
    \multicolumn{3}{c}{\textrm{Camera settings}}\\\hline
    \object{BPM~16274} & 0125320401 & 33\,728 & EPN EFF & MOS1 FF & MOS2 FF \\ 
                       & 0125320501 &    7845 & EPN FF  & MOS1 FF & MOS2 FF \\
                       & 0133120301 & 12\,022 &EPN FF& MOS1 FF &  MOS2 FF\\
                       & 0133120401 & 13\,707 &EPN FF& MOS1 FF &  MOS2 FF\\
                       & 0125320701 & 45\,951 &EPN FF& MOS1 FF &  MOS2 FF\\
                       & 0153950101 &    5156 &EPN FF& MOS1 FF &  MOS2 FF\\
    \hline
    CFRS 3h & 0041170101 & 51\,724 & EPN EFF & MOS1 FF & MOS2 FF \\
    \hline
    \object{RX J0505.3$-$2849} & 0111160201& 49\,616 &EPN EFF & MOS1 FF & MOS2 FF
    \\  
    \hline
    \object{RBS~0864} & 0108670101 & 56\,459 & EPN FF & MOS1 FF & MOS2 FF \\
    \hline
    \object{QSO B0130$-$403} &0112630201& 37\,870 &EPN FF & MOS1 FF & MOS2 FF \\ 
    \hline
    \object{BR~1033$-$0327} &0150870401& 31\,418 &EPN FF & MOS1 FF & MOS2 FF \\ 
    \hline 
    \object{SDSS J104433.04$-$012502.2} & 0125300101 & 62\,310 & EPN FF & MOS1 FF & MOS2 FF \\
    \hline
    \object{MS1054.4$-$0321} &0094800101&41\,021 &EPN FF & MOS1 FF & MOS2 FF \\ 
    \hline
    \object{HE~1104$-$1805} &0112630101&36\,428 &EPN FF & MOS1 FF & MOS2 FF \\
    \hline
    \object{PG~1115+080} & 0082340101&63\,358 &EPN FF & MOS1 FF & MOS2 FF \\ 
                         & 0203560201&86\,649 &EPN FF & MOS1 FF & MOS2 FF  \\
                         & 0203560401&86\,515 &EPN FF & MOS1 FF & MOS2 FF \\
    \hline
    \object{CD~$-$33 07795} &0112880101&29\,921 &EPN FF & MOS1 FF & MOS2 FF \\
    \hline
    \object{T Leo} & 0111970701&12\,866 &EPN FF & MOS1 SW3 & MOS2 SW2  \\
    \hline
    \object{IRAS~12112+0305} &0081340801&23\,206 &EPN FF & MOS1 FF & MOS2 FF \\ 
    \hline
    \object{LBQS~1228+1116} &0145800101&107\,002&EPN FF & MOS1 FF & MOS2 FF \\ 
    \hline
    \object{NGC~4666} &0110980201&58\,237 &EPN EFF & MOS1 FF & MOS2 FF \\
    \hline
    \object{QSO\,B1246$-$057} & 0060370201 & 41\,273 & EPN FF & MOS1 FF & MOS2 FF\\  
    \hline
    Field 864-1 & 0111281001&10\,377 &EPN EFF & MOS1 FF & MOS2 FF \\ 
    \hline
    Field 864-2 & 0111282401 & 7077 & EPN EFF & MOS1 FF & MOS2 FF\\
    \hline
    Field 864-4 & 0111281301 & 14\,541 & EPN EFF & MOS1 FF & MOS2 FF\\
    \hline
    Field 864-5 & 0111281401 & 8643 & EPN EFF & MOS1 FF & MOS2 FF\\
    \hline
    Field 864-6 &0111281501&8650  &EPN EFF & MOS1 FF & MOS2 FF \\ 
    \hline
    Field 864-9 & 0111282501&8623  &EPN EFF & MOS1 FF & MOS2 FF \\
    \hline
    \object{A~1882} &0145480101&23\,567 &EPN FF & MOS1 FF & MOS2 FF \\
    \hline
    \object{MKW~9} &0091140401&45\,414 &EPN EFF & MOS1 FF & MOS2 FF \\
    \hline
    \object{LBQS~2212$-$1759} & 0106660101 &  60\,508 & EPN FF &  MOS1 FF &  MOS2 FF \\ 
                              & 0106660201 &  53\,769 & EPN FF & MOS1 FF & MOS2 FF\\ 
                              & 0106660401 &  35\,114 & ---       & MOS1 FF &  MOS2 FF\\  
                              & 0106660501 &  11\,568 & EPN FF &  MOS1 FF &  MOS2 FF \\   
                              & 0106660601 & 110\,168 & EPN FF & MOS1 FF & MOS2 FF\\ 
    \hline
    \object{NGC~7252} & 0049340201 & 28\,359  & EPN FF & MOS1 FF & MOS2 FF \\ 
    \hline
    \object{PHL~5200} & 0100440101 & 46\,681  & EPN FF & MOS1 FF & MOS2 FF\\
    \hline
    \end{tabular}
    \label{tab:xray-observations}
\end{table*}

\end{document}

%% file: 7281tab1.tex
\begin{table*}[t]
  \centering
  \caption{Effective area and number density of galaxies of WFI fields in the XFS
    used in the cluster search.} 
  \begin{tabular}{l%
      >{$}r<{$}@{:}>{$}c<{$}@{:}>{$}l<{$}
      >{$}r<{$}@{:}>{$}c<{$}@{:}>{$}l<{$}
      >{$}r<{$}
      >{$}r<{$}
      >{$}r<{$}
      >{$}r<{$}
      >{$}r<{$}
    }
    \hline\hline
    Field & 
    \multicolumn{3}{c}{$\alpha$} & 
    \multicolumn{3}{c}{$\delta$} &
    \multicolumn{1}{c}{$m_\mathrm{lim}$}&
    \multicolumn{1}{c}{Seeing} &
    \multicolumn{1}{c}{Area} & 
    \multicolumn{1}{c}{Number Density (WL)} &
    \multicolumn{1}{c}{Number Density (MF)} \\ 
    & 
    \multicolumn{3}{c}{(J$2000.0$)} & 
    \multicolumn{3}{c}{(J$2000.0$)} & 
    \multicolumn{1}{c}{(mag)} &
    \multicolumn{1}{c}{(arcsec)} &
    \multicolumn{1}{c}{(arcmin$^2$)} &
    \multicolumn{1}{c}{(arcmin$^{-2}$)} &
    \multicolumn{1}{c}{(arcmin$^{-2}$)}  \\
    \hline
    BPM~16274                   & 00&50&03.2 & -52&08&17 & 25.38 & 0.88 & 1011 & 15.6 & 21.6 \\
    QSO\,B0130$-$403            & 01&33&01.9 & -40&06&28 & 24.94 & 1.01 &  925 &  7.8 & 12.7 \\ 
    CFRS 3h                     & 03&02&39.2 & +00&07&31 & 25.46 & 0.85 &  965 & 12.1 & 17.0 \\
    RX~J0505.3$-$2849          & 05&05&20.0 & -28&49&05 & 25.58 & 0.83 &  981 & 18.2 & 25.9 \\
    RBS~0864-N                  & 10&21&03.8 & +04&26&23 & 25.53 & 1.01 &  948 & 13.0 & 20.2 \\
    QSO~B1033$-$033              & 10&36&23.7 & -03&43&20 & 25.59 & 0.78 & 1037 & 16.8 & 22.0 \\
    SDSS~J104433.04$-$012502.2 & 10&44&33.0 & -01&25&02 & 25.44 & 0.73 &  997 & 16.9 & 22.6 \\
    MS~1054.4$-$0321            & 10&56&60.0 & -03&37&27 & 25.35 & 0.69 & 1022 & 17.7 & 23.5 \\
    HE~1104$-$1805              & 11&06&33.0 & -18&21&24 & 25.47 & 0.83 & 1030 & 13.4 & 22.4 \\
    PG~1115+080                 & 11&18&17.0 & +07&45&59 & 25.48 & 0.92 & 1051 & 12.3 & 19.6 \\
    CD~$-$33~07795              & 11&29&27.2 & -34&19&55 & 25.04 & 0.91 &  927 &  6.1 & 19.1 \\
    T~Leo                       & 11&38&27.1 & +03&22&10 & 25.47 & 0.74 & 1005 & 18.6 & 23.9 \\
    IRAS~12112+0305             & 12&13&46.1 & +02&48&41 & 25.51 & 0.88 & 1028 & 13.9 & 21.9 \\
    LBQS~1228+1116              & 12&30&54.1 & +11&00&11 & 25.09 & 1.01 & 1014 &  8.4 & 14.3 \\
    NGC~4666                    & 12&45&08.9 & -00&27&38 & 25.43 & 0.87 & 1074 & 10.7 & 17.8 \\
    QSO~B1246$-$057            & 12&49&13.9 & -05&59&19 & 25.21 & 0.80 &  869 & 13.0 & 19.6 \\
    Field~864-1                 & 13&41&22.4 & +00&23&52 & 25.45 & 0.88 & 1058 & 16.8 & 23.9 \\
    Field~864-9                 & 13&44&36.0 & -00&24&00 & 25.56 & 0.84 & 1045 & 16.9 & 25.6 \\
    A~1882                      & 14&14&39.9 & -00&19&07 & 25.73 & 0.72 & 1019 & 18.6 & 29.5 \\
    MKW~9                       & 15&32&29.3 & +04&40&54 & 25.56 & 0.91 & 1032 & 10.7 & 19.6 \\
    LBQS~2212$-$1759            & 22&15&31.7 & -17&44&05 & 25.52 & 0.99 &  970 & 12.6 & 19.9 \\
    NGC~7252                    & 22&20&44.8 & -24&40&42 & 25.57 & 0.70 &  997 & 20.3 & 26.4 \\
    PHL~5200                    & 22&28&30.4 & -05&18&55 & 24.92 & 1.06 & 1068 &  7.8 & 11.7 \\
    \hline
    \multicolumn{9}{l}{Total/average} & \textbf{23073} & \textbf{14.1} & \textbf{20.9}\\
    \hline
  \end{tabular}
\label{tab:optical_fields}
\end{table*}